\documentclass[aps,twocolumn,floats,prd,nofootinbib,10pt,longbibliography,superscriptaddress]{revtex4-2}

\usepackage[dvips]{graphicx} %
\usepackage{graphicx,amsmath,amsfonts,amssymb,slashed,float,hyperref}
\usepackage[normalem]{ulem}
\usepackage{bbold,wasysym}
\usepackage{epsfig}
\usepackage{array,multirow}
\usepackage[utf8]{inputenc}

\usepackage[usenames,dvipsnames]{xcolor} 

\usepackage{soul}

\begin{document}

\title{Early Dark Energy beyond slow-roll: implications for cosmic tensions}
\author{Ravi kumar Sharma}
\affiliation{Indian Institute of Astrophysics, Bengaluru, Karnataka 560034, India}
\author{Subinoy Das}
\affiliation{Indian Institute of Astrophysics, Bengaluru, Karnataka 560034, India}
\author{Vivian Poulin }
\affiliation{Laboratoire Univers and Particules de Montpellier(LUPM), CNRS, Universit\'e de Montpellier (UMR-5299)}

\begin{abstract}
 
In this work, we explore the possibility that Early Dark Energy (EDE) is dynamical in nature and study its effect on cosmological observables. We introduce a parameterization of the equation of state allowing for an equation of state $w$ differing considerably from cosmological constant (cc, $w={-1}$) and vary both the initial $w_i$ as well final $w_f$  equation of state of the EDE fluid. This idea is motivated by the fact that in  many models of EDE, the scalar field may have some kinetic energy when it starts to behave like EDE before the CMB decoupling. We find that the present data have a mild preference for non-cc early dark energy $( w_i= -0.78)$ using Planck+BAO+Pantheon+S$H_0$ES data sets, leading to $\Delta \chi^2_{\rm min}$ improvement of -2.5 at the expense of one more parameter. However, $w_i$ is only weakly constrained, with $w_i < -0.56$ at $1\sigma$. We argue that allowing for $w_i\neq -1$ can play a role in decreasing the $\sigma_8$ parameter. Yet, in practice the decrease is only $\sim0.4\sigma$ and $\sigma_8$ is still larger than weak lensing measurements. We conclude that while promising, a dynamical EDE cannot resolve both $H_0$ and $\sigma_8$ tensions simultaneously. 
\end{abstract}

\maketitle

\section{Introduction}
The cosmological model with a cosmological constant and cold dark matter dubbed `$\Lambda$CDM'  has been the most successful candidate to describe our universe. It is consistent with almost all cosmological observations, but recently a few discrepancies between the model predictions and direct measurements of a few observables have arisen. Measurement of the expansion rate of the Universe is one of the most discussed mismatches over the last decade. The latest measurements of the Cosmic Microwave Background (CMB) by the Planck satellite when analyzed under the $\Lambda$CDM model predict a value of the Hubble parameter $H_0=67.36\pm 0.54 \rm {km/s/Mpc}$ \cite{pl18}.  Beyond Planck, data that are calibrated using pre-recombination information, like Baryon Acoustic Oscillations(BAO)\cite{DES:2017txv} and Big Bang nucleosynthesis(BBN)\cite{Cooke:2016rky} is consistent with the lower value of Hubble parameter inferred from CMB.
On the other hand, the Supernovae H0 for the equation of state (S$H_0$ES) team has measured the value of the Hubble parameter $H_0=73.04\pm 1.04 \rm {km/s/Mpc}$  by building a distance ladder to supernovae of type 1a (SN1a) \cite{Riess:2019cxk,Riess:2020fzl,Riess:2021jrx}. Other local/direct measurements are consistent with a higher value of the Hubble parameter, although not at the same tension level as S$H_0$ES (see Ref.~\cite{Abdalla:2022yfr} for a review). With the increasing precision of experiments, and more and more data available, this tension has gained more significance and drawn more attention from the cosmology community \cite{Schoneberg:2021qvd, DiValentino:2020zio, DiValentino:2021izs}.\\

There also exists a comparatively milder tension in the measurement of local growth parameterized by $S_8=\sigma_8 \sqrt{\frac{\Omega_{M}}{0.3}}$, where $\Omega_{M}$ is the total matter density today and $\sigma_8$ is root mean square of matter fluctuations at the scale of 8 Mpc$/h$. 
The Planck 2018 CMB measurements using $\Lambda$CDM model infers the value of $ S_=0.832\pm {0.013}$. On the other hand, observations of galaxies through weak lensing by CFHTLenS collaboration have indicated that the $\Lambda$CDM model predicts a $S_8$ value that is larger than the direct measurement at the $2\sigma$ level \cite{Heymans:2013fya,MacCrann:2014wfa}.  
This tension has gained more significance with various data sets such as the KiDS/Viking data \cite{Hildebrandt:2018yau,Joudaki:2019pmv} and DES data \cite{Abbott:2017wau,DES:2021wwk}. Recently, the combination of KiDS/Viking and SDSS data has established $3\sigma$ tension with $S_8=0.766^{+0.02}_{-0.014}$ \cite{Heymans:2020gsg}, although the combination of KiDS and DES indicates a slightly lower significance of the tension \cite{Kilo-DegreeSurvey:2023gfr}.

No studies have found obvious errors and systemic in data that could explain the $H_0$ tension, such that the possibility of new physics has gained a lot of attention. It is becoming evident  from recent works \cite{Knox:2019rjx,Schoneberg:2021qvd} that the pre-recombination era  is the most likely epoch to contain hidden new physics which may solve the Hubble tension by reducing the CMB sound horizon. 
Early dark energy, originally introduced in Refs.~\cite{Karwal:2016vyq,Poulin:2018cxd,Lin:2019qug}, invokes a scalar field frozen until matter-radiation equality, that suddenly becomes dynamical and dilutes faster than radiation. Although EDE can bring down the $H_0$ tension significantly,  it suffers from a few major challenges (see Refs.~\cite{Kamionkowski:2022pkx,Poulin:2023lkg} for discussion). First, generic to models that resolve the Hubble tension by changing the pre-recombination era, the EDE cosmology has a small-scale power increase in the matter power spectra, that tends to slightly worsen the $S_8$ tension \cite{Hill:2020osr,Vagnozzi:2021gjh}. Second, the EDE model suffers a problem of coincidence, whereby the EDE must become dynamical at a special time, namely matter-radiation equality, that raises questions of fine-tuning in the model (see e.g. Refs.~\cite{Gogoi:2020qif,Niedermann:2021ijp,Lin:2022phm} for  studies in this context). At the same time, at late time the existence of the present dark energy is still a mystery and faces a similar challenge of fine-tuning. Dynamical dark energy is one of the most well-studied candidates  as an alternative to cosmological constant to explain the fine-tuning problem.\\

There already have been a few physical models where early dark energy appears to be naturally present at matter radiation equality \cite{Gogoi:2020qif,CarrilloGonzalez:2023lma,Lin:2022phm}. In some physical models, one can in fact have a dynamical evolution for early dark energy even prior to its fast dilution. For example in Ref.~\cite{Gogoi:2020qif}, authors discuss a model where a neutrino-like particle is in interaction with a scalar field. Depending on the shape of the potential of this scalar field, the equation of state of EDE can change, and it is shown that for a $\phi^2$ potential, the {\it initial} equation of state is $w_i=-\frac{1}{3}$.

 In this paper, motivated by this idea, we explore whether a dynamical EDE model may be preferred over a purely frozen (i.e., cosmological constant-like) behavior. If EDE also happens to be dynamical, one might get a hint that perhaps `Nature' has the same mechanism of turning on dark energy at different epochs of the universe - may it be inflation or EDE or present DE. Interestingly, we indeed find that  up-to-date cosmological data sets may indeed prefer an equation of state $w_i \neq -1$.
The plan of the paper is as follows: In section \ref{sec:background} we present our phenomenological modeling of the EDE component at the background and perturbations level. We discuss the impact of the initial equation of state (before the fluid dilutes) on observable in section \ref{sec:effects}. In section \ref{sec:datasets},  we present our main analysis setup and methodology, while our results are discussed in section \ref{sec:results}. We finally conclude in section \ref{sec:concl}.

\section{Background and Perturbation equations}\label{sec:background}
We consider a homogeneous isotropic and  flat universe described by the Friedmann-Lemaitre-Robertson-Walker (FLRW) metric and filled with the usual species that compose the $\Lambda$CDM model (photons, baryons, neutrinos, cold dark matter, and dark energy). To implement EDE, we consider an extra component in the form of a generalized fluid description, which requires us to specify its equation of state $w$, sound speed $c_s^2$, and eventually, its anisotropic stress which we take to be zero as valid for a scalar field (see Ref.~\cite{Sabla:2022xzj} for a discussion about the role of anisotropic stress with EDE). The equation of state is parameterized as follows:
\begin{equation}
w_{\rm EDE}(a) = {\frac{{w_f}-{w_i}}{\left[{1+{\left(\frac{a_c}{a}\right)}^{3\times(w_f-w_i)}}\right]}}-w_i
\label{eos}
\end{equation}

where $w_i$ and $w_f$ are the initial and final equations of states parameters respectively and $a_c$ is the scale factor at the time of transition, while $p$ is the parameter controlling the width of the transition. The background  energy density of the early dark energy component evolves as follows
\begin{equation}
    \rho_{\rm EDE}(a)=\rho_{\rm EDE}(1) \times \exp\left(3\int_{1}^{a}  (1+w_{\rm EDE}(a))da\right) \,.
\end{equation}
To describe perturbations in the fluid, we make use of  the generalized dark matter formalism \cite{Hu:1998kj}. The perturbation
equations (Euler and Continuity) in the synchronous gauge are given by:


\begin{eqnarray}\label{eq:Cont}
\frac{d}{d\eta} \left(\frac{\delta_{\rm EDE}}{1+w_{\rm EDE}}\right) &=& -\left(\theta_{\rm EDE}+h'\right) \nonumber \\
&& - 3 \textit{H}(c_s^2-c_a^2)  \left(\frac{\delta_{\rm EDE}}{1+w_{\rm EDE}} + 3 \textit{H} \frac{\theta_{\rm EDE}}{k^2}\right)\,,\nonumber
\end{eqnarray}

\begin{equation}
  \frac{d}{d\eta}\left(\theta_{\rm EDE}\right) = -\textit{H} \left(1-3 c_s^2\right) \theta_{\rm EDE} + c_s^2 k^2 \frac{\delta_{\rm EDE}}{1+w_{\rm EDE}}\,.
\label{eq:Euler}  
\end{equation}

Here $\delta_{\rm EDE}$ is the density perturbation and $\theta_{\rm EDE}$ is the velocity divergence of the EDE fluid. The sound speed of the EDE fluid relates the density and pressure perturbations as $c_s^2=\frac{\delta P}{\delta \rho}$, $k$ is the comoving wave-number, $\textit{H}$ the conformal Hubble parameter, $c_a^2$ is the adiabatic sound speed defined as 
\begin{equation}
    c_a^2 = \frac{\rho'_{\rm EDE}}{P'_{\rm EDE}} = w_{\rm EDE} - \frac{1}{3} \frac{dw_{\rm EDE}/d\ln a}{1+w_{\rm EDE}}\,.
\end{equation}

Using equation \ref{eos}, we find that 
$$
c_a^2=
\begin{cases}
     w_i & a\ll a_c \,,\\
      w_f &a\gg a_c \,.
\end{cases}
$$
The sound speed $c_{s}^2$ is set as follows
$$
c_s^2=
\begin{cases}
     1 & a < a_c \,,\\
      w_f &a >= a_c \,,
\end{cases}
$$
but we find that the results are not strongly sensitive to the way in which we parameterize $c_s^2$, when we impose $c_s^2 = w_f$ in the decaying phase.

To find the initial conditions, we assume that the system starts in the radiation-dominated era so $\textit{H}=\frac{1}{\eta}$. If the energy density of early dark energy at early times is negligible, the solution for the metric perturbation $h$ will not change and is given by $h=\frac{(k\eta)^2}{2}$. For super–Hubble
mode  $k\eta \ll 1$, equations \ref{eq:Cont} and  \ref{eq:Euler} reduced to 

$$
\frac{d}{d\eta} \left(\frac{\delta_{\rm EDE}}{1+w_{\rm EDE}}\right) = \frac{k^2\eta}{2}-3\frac{1}{\eta}(c_s^2-c_a^2)\left(\frac{\delta_{\rm EDE}}{1+w_{\rm EDE}} + 3 \frac{1}{\eta} \frac{\theta_{\rm EDE}}{k^2}\right)$$

$$
\frac{d}{d\eta}\left(\theta_{\rm EDE}\right) = -\frac{1}{\eta} \left(1-3 c_s^2\right) \theta_{\rm EDE} + c_s^2 k^2 \frac{\delta_{\rm EDE}}{1+w_{\rm EDE}}\,.
$$

These can be solved in power of $(k\eta)^2$, and we get the initial condition  

\begin{eqnarray}
\frac{\delta_{\rm EDE}}{1+w_{\rm EDE}} &=& -\frac{(4-3 c_s^2)/2}{8+6 c_s^2 - 12 c_a^2} (k \eta)^2, \label{eq:deltaIC}\\
\theta_{\rm EDE} &=& -\frac{c_s^2/2 }{8+6 c_s^2 - 12 c_a^2}k (k \eta)^3\,.\label{eq:thetaIC}
\end{eqnarray}
These equations and the initial conditions have been implemented in a modified version of the Boltzmann code {\sf CLASS} \cite{Lesgourgues:2011re,Blas:2011rf}.

\begin{figure*}
\centering
\includegraphics[width=0.45 \textwidth]{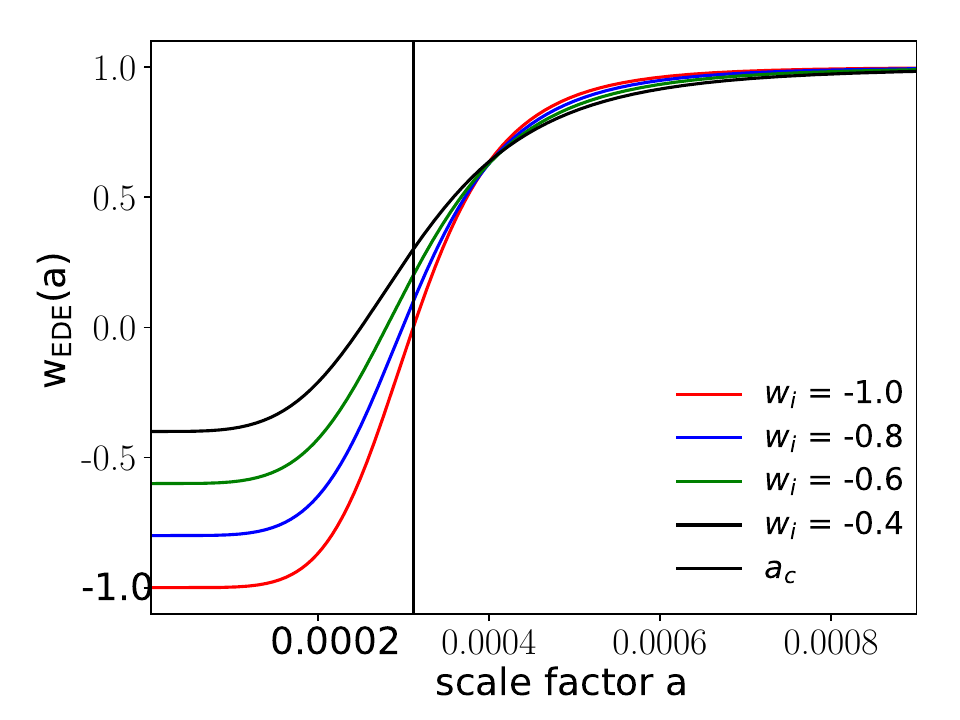}
\includegraphics[width=0.45 \textwidth]{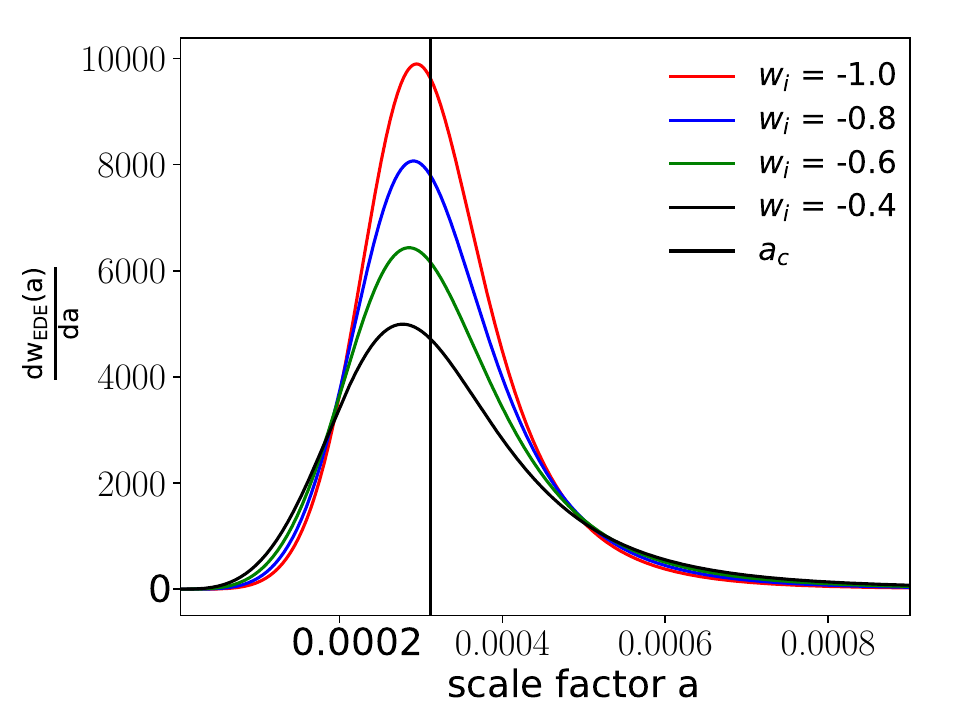}
\caption{(a) Effect of varying $w_i$ on the equation of state. (b) Effect of varying $w_i$ on the slope. }
\label{width_1}
\end{figure*}

\begin{figure}
\centering
\includegraphics[scale=0.45]{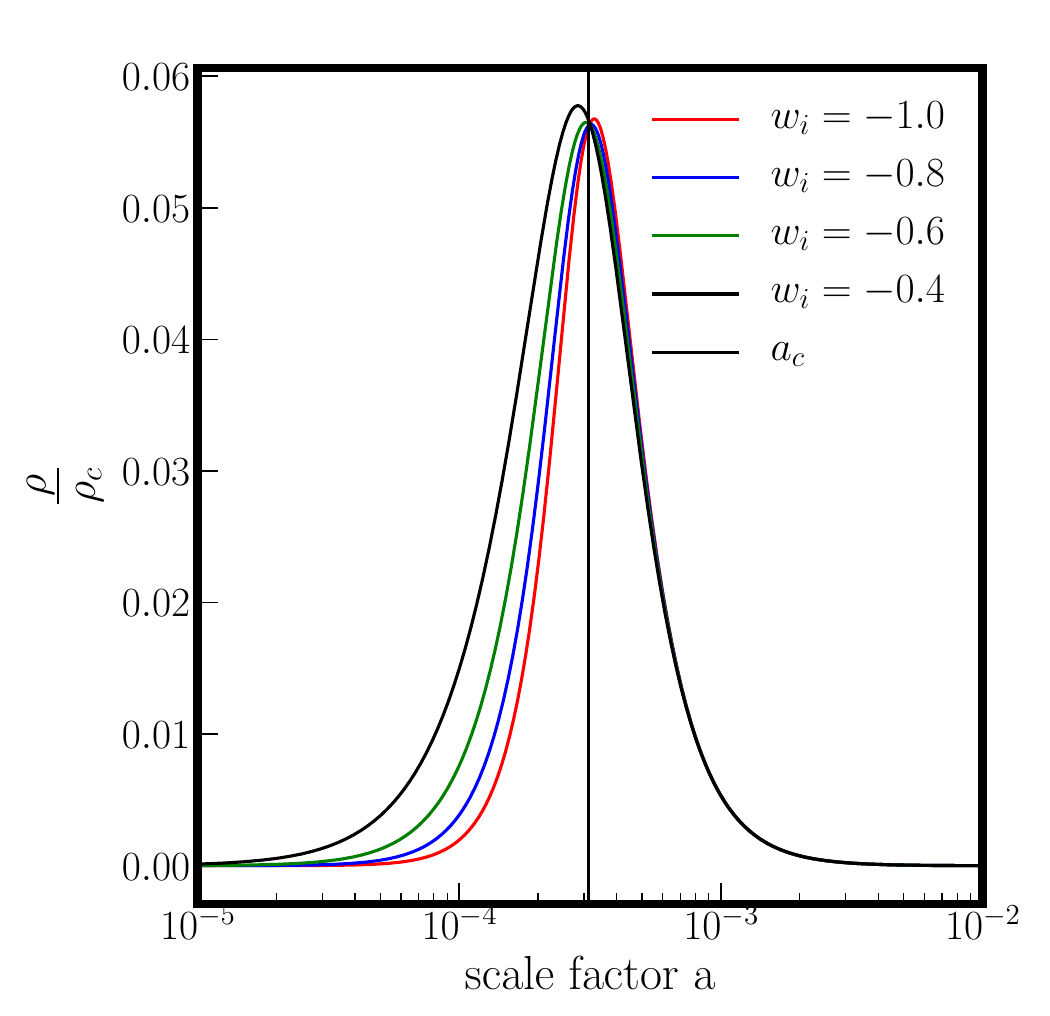}
\caption{Effect of varying $w_i$ on the evolution of EDE density fraction.}
\label{width_3}
\end{figure}

\section{Effect of changing the initial equation of state $w_i$}
\label{sec:effects}

\subsection{Impact on the background and perturbation dynamics}
To investigate the effect of changing on  background and perturbations quantities, we vary $w_i$ for a set of values as $w_i=-1,-0.8,-0.6,-0.4$. The effect of varying $w_i$ on $w_{\rm EDE}(a)$ and $\frac{dw_{\rm EDE}(a)}{da}$ is illustrated on Fig.~\ref{width_1}.

\textit{\textbf{Impact of $w_i$ on EDE background evolution}}:
Firstly let us investigate the effect on the evolution of the EDE background energy density. This is shown in figure \ref{width_3}. One can clearly notice that when we increase $w_i$, the energy density spreads over time.  The transition is smoother,  the EDE component stays longer and the EDE component starts influencing the expansion rate from earlier redshift.  That results in a smaller value of sound horizon and a large value of the Hubble parameter today.\\

\begin{figure*}[t!]
\centering
\includegraphics[width= 1.9\columnwidth]{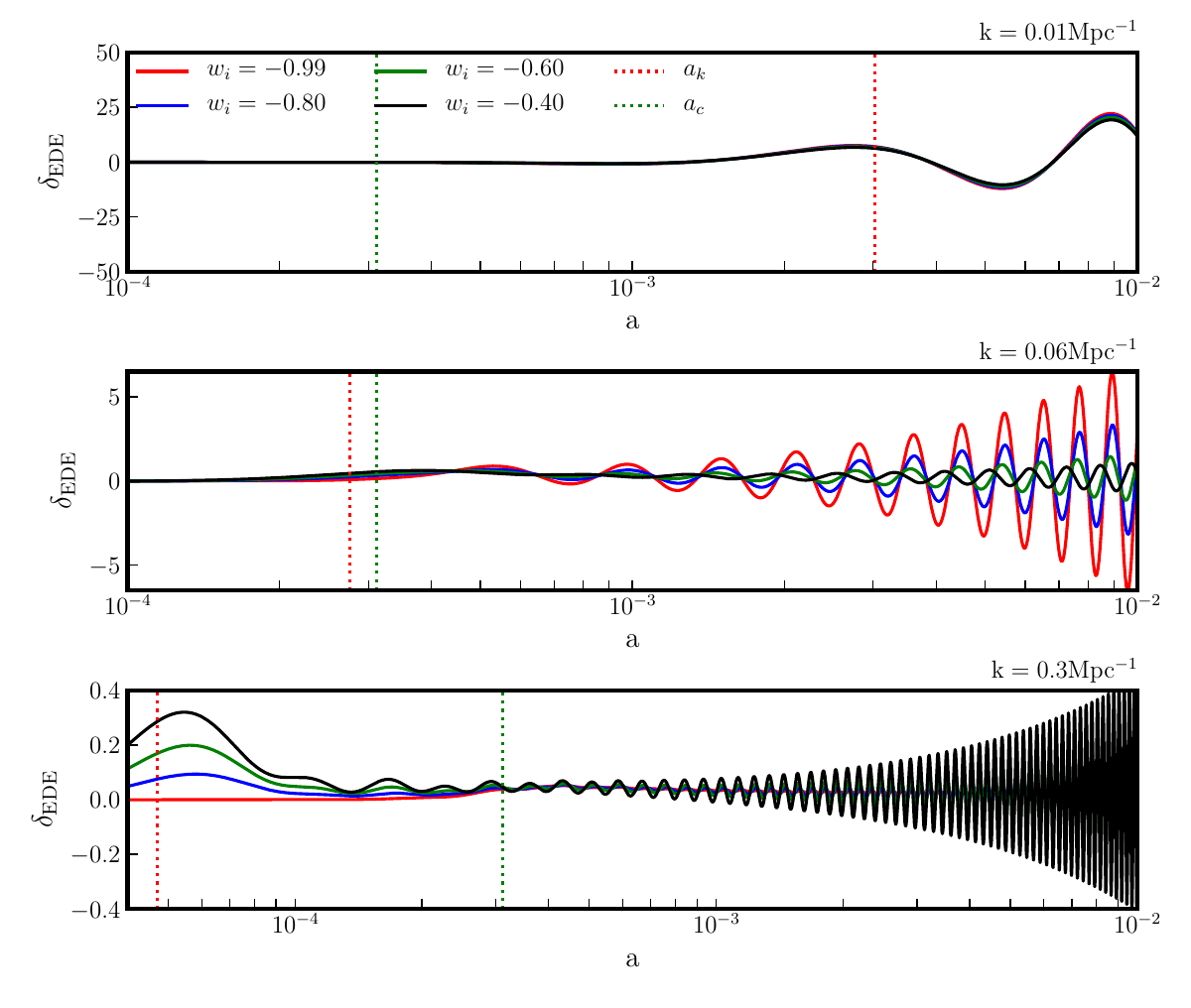}
\caption{Effect of varying $w_i$ on the evolution of EDE perturbations for three modes  $k=0.01,0.06,0.3$Mpc$^{-1}$. }
\label{delta_ede}
\end{figure*}

\begin{figure*}
\centering
\includegraphics[width= 1.9\columnwidth]{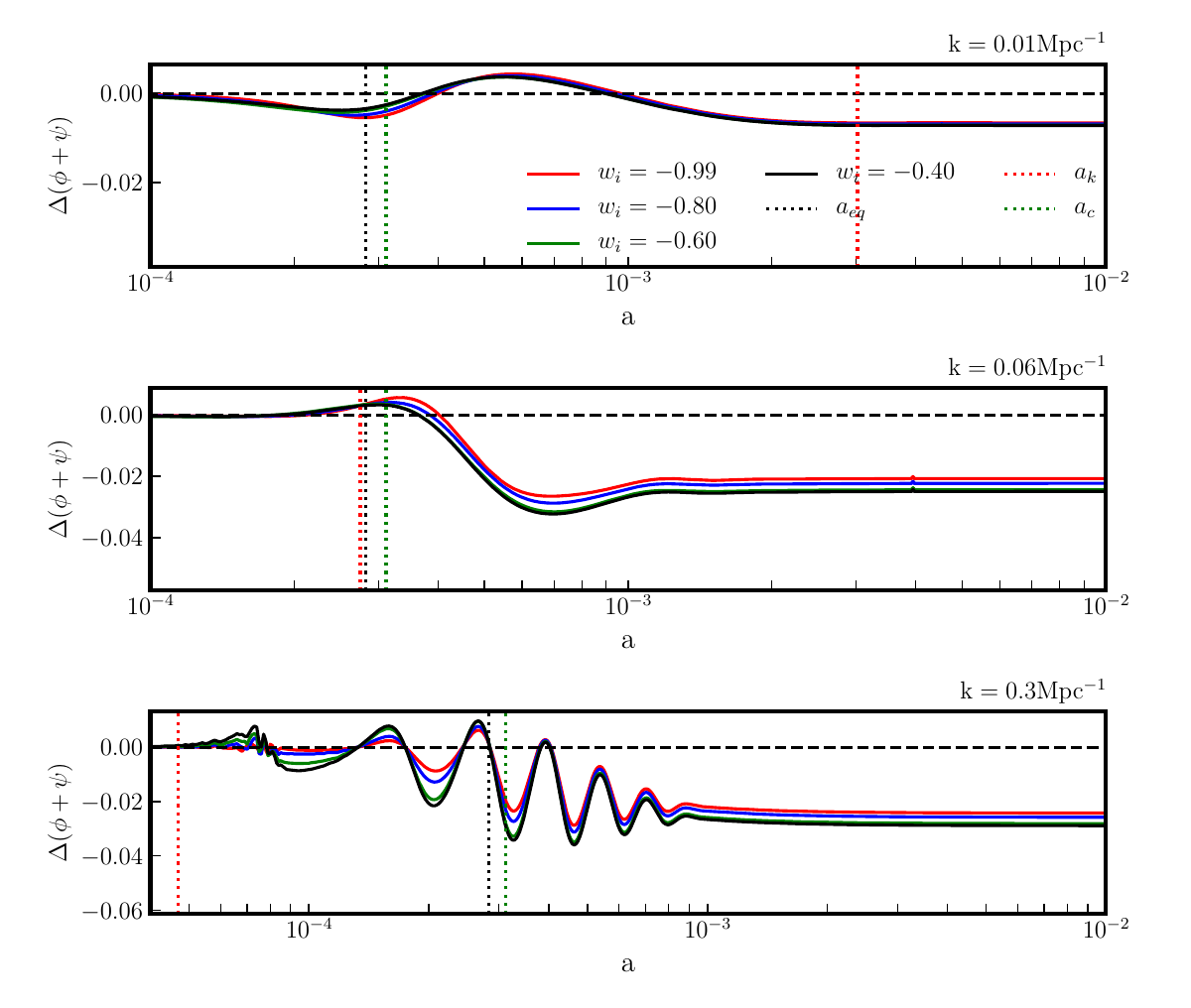}
\caption{Effect of varying $w_i$ on the evolution of weyl potential. We show the difference with respect to the $\Lambda$CDM model with identical cosmological parameters. }
\label{weyl_potential}
\end{figure*}

We now turn to describe the effect of $w_i$ on perturbations of early dark energy as well as how it affects the matter component.\\

\textit{\textbf{Impact of $w_i$  on EDE density perturbations}}:\\
In figure \ref{delta_ede} we plot the effect of varying $w_i$ on EDE density fluctuations, for $k=0.01,0.06,0.3$Mpc$^{-1}$, i.e. modes that enter the horizon  after, around and before $a_c$ respectively.  We also show horizon crossing for each mode defined as $ k\tau(a_k) \equiv 2\pi$ and the value of $a_c$ as red and green dotted vertical line respectively. 
The equation which governs EDE perturbations can be obtained by reducing Eqs. \ref{eq:Cont} and \ref{eq:Euler} to a second-order differential equation,
\begin{eqnarray}
\frac{d^2}{d\eta^2}\left(\frac{\delta_{\rm EDE}}{1+w_{\rm EDE}}\right)& + & k^2 c_s^2 \frac{\delta_{\rm EDE}}{1+w_{\rm EDE}} \\
& + & (1-3c_a^2) \frac{a'}{a} \frac{d}{d\eta}\left(\frac{\delta_{\rm EDE}}{1+w_{\rm EDE}}\right) =0\,.\nonumber
\label{eq:delta_ede}
\end{eqnarray}
This equation is that of a damped simple harmonic oscillator. The  solution will be oscillatory either decreasing in amplitude or increasing in amplitude depending on the damping terms sign. The frequency of oscillations depends on '$k^2 c_s^2$', and 
The term $1-3 c_a^2$ can be either negative or positive, acting either as a driving force of a friction term (assuming $w_i < -\frac{1}{3}$):

$$
1-3 *c_a^2=
\begin{cases}
     1-3*w_i>0 & a\ll a_c \\
      1-3*w_f<0 &a\gg a_c \,.
\end{cases}
\label{eqn:damp}
$$

The effect of $w_i$ on the growth of the EDE density fluctuations for different $k$ modes can be understood as follows: 

\begin{itemize}
   \item For modes which enter the horizon well before $a_c$, i.e. for $a_k\ll a_c$: Before horizon crossing, the growth of $\delta_{\rm EDE}$ is given by the initial condition [Eq.~(\ref{eq:deltaIC})], which shows that it grows like $(k\tau)^2$ and proportionally to $1+w_{\rm EDE}$. Consequently, models with larger $w_i$ grow faster, as is particularly visible for the mode with $k=0.3$Mpc$^{-1}$, which crosses the horizon earlier.
   In the region $a_{k}<a<a_{c}$, the equation of state is $w_i<0$. The solution of $\delta_{\rm EDE}$ is oscillatory but decreasing in amplitude (see equation \ref{eq:delta_ede}). At the time of $a_c$, high $w_i$ models have a higher amplitude. Finally, in the region $a>a_{c}$, the solution remains oscillatory, but the damping factor switches sign, acting as a driving force and leading to oscillations whose amplitude is larger for modes that had a larger amplitude at $a_c$, i.e., modes with larger $w_i$.

   \item  For modes which enter the horizon well after $a_c$, i.e. that verifies $ a_{k} \gg a_c$ : In that case, all modes have a similar evolution since they are frozen (i.e.  $(k\tau)^2\ll 1$) when the fields become dynamical. Around horizon crossing, all modes grow proportional to $1+w_{\rm ede}$ as given by [Eq.~(\ref{eq:deltaIC})] while after horizon crossing they oscillate with increasing amplitude as larger modes do.

   \item  For modes which enter the horizon around $a_c$, i.e. $a_k\sim a_c$: Before and around horizon crossing, the growth of $\delta_{\rm EDE}$ is still given by the initial condition [Eq.~(\ref{eq:deltaIC})], and modes oscillate with increasing amplitude after horizon crossing. However, $w_{\rm EDE}$ evolves in time as the modes enter the horizon, and can even become positive around $a_k$. This gives a non-trivial time-evolution to the damping term, such that modes with larger amplitude at $a_k$ now have a {\it smaller} amplitude of oscillations at $a>a_k$. For instance, the model with $w_i = -0.4$, shows a much lower amplitude of oscillations than the model with $w_i = -0.99$ at this scale, while it shows a much larger oscillation amplitude at larger $k$. We stress that this is a specific consequence of our choice of parameterization of $w_{\rm EDE}(a)$, rather than the effect of $w_i$ per se.
   
   \end{itemize}


\textit{\textbf{Impact on the Weyl potential Evolution}}:\\
To understand the impact of $w_i$ on the CMB power spectra, it is instructive to plot the combination of $-(\Psi+\Phi)$ known as Weyl potential and determined as \cite{Poulin:2023lkg} 
\begin{equation}
    \Phi = -\frac{3}{4k^2} \left(\frac{a'}{a}\right)^2 \left(2\delta + \sum_i [1 + w_i]\left[\frac{6(a'/a)\theta_i}{k^2} + 3\sigma_i\right]\right)\,.
\end{equation}

We show in Fig.~\ref{weyl_potential} the effect of $w_i$ on the evolution of the Weyl potential for different $k$ modes (identical to the previous figure), normalized to the standard $\Lambda$CDM case. The impact on the Weyl potential can be explained through a combination of background and perturbation effects which have been discussed above. \\

\begin{figure*}
\centering
\includegraphics[width=0.49 \textwidth]{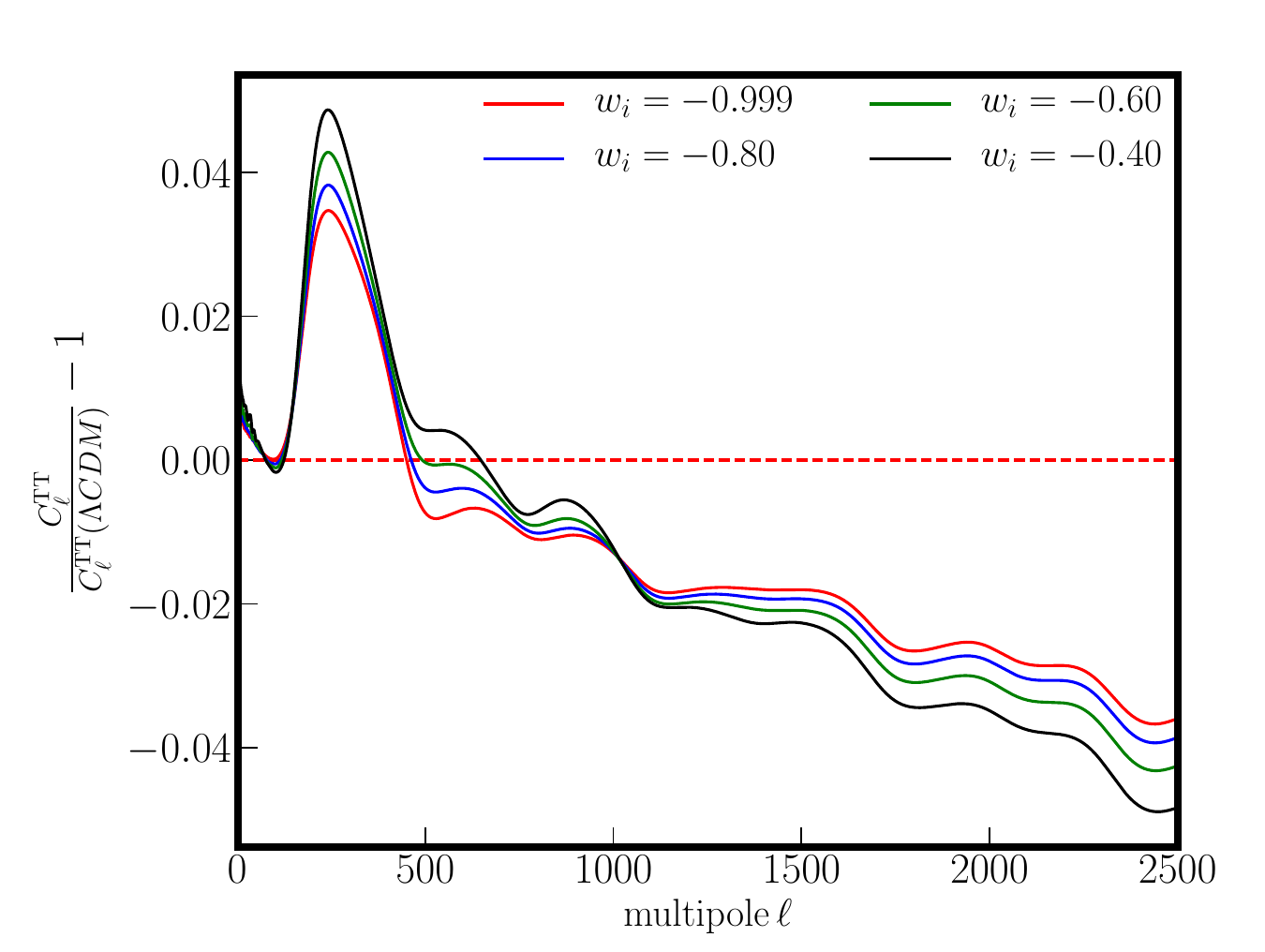}
\includegraphics[width=0.49\textwidth]{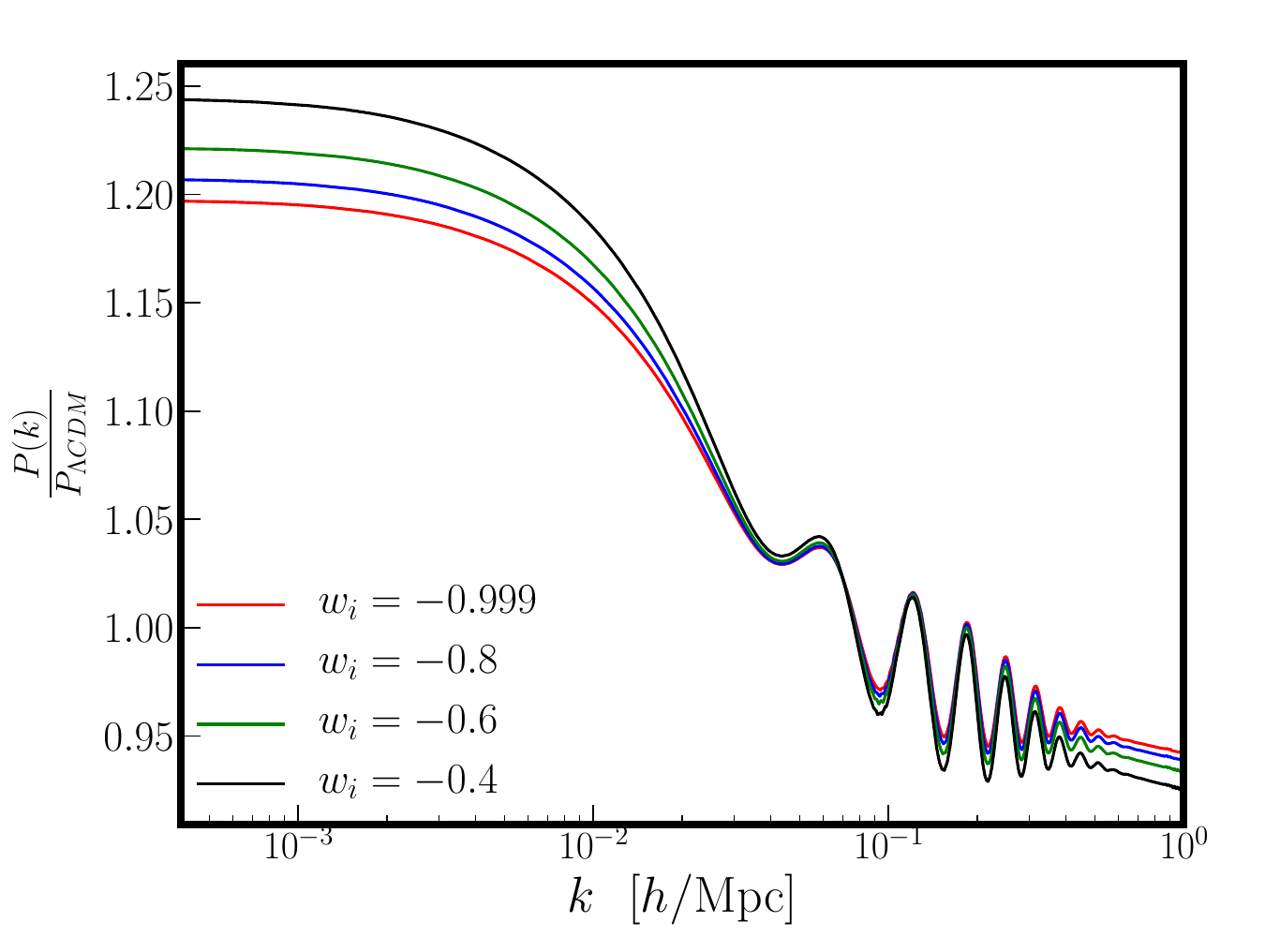}
\caption{Effect of varying $w_i$ on CMB TT power spectra (left panel) and 
 matter power spectra (right panel).}
\label{ede_cl_pk}
 \end{figure*}

\begin{itemize}
    \item At $a\gg a_c$ the Weyl potential is suppressed due to the presence of EDE, which contributes to the Hubble rate but does not cluster. For modes that are within the horizon before $a_c$, the larger $w_i$, the longer the EDE phase lasts, and the more the Weyl potential is suppressed. For modes entering the horizon much later, the suppression is identical since the modes are mostly sensitive to EDE after $a_c$, when all models are identical.     
    
    \item At $a\approx a_c$, there are  visible residual oscillations (decaying in amplitude) that are due to EDE perturbations. The frequency of oscillations is larger for larger $k$ modes, as the EDE oscillations have a frequency set by $(kc_s)^2$.  For modes that are within the horizon before $a_c$ ($k=0.3$Mpc$^{-1}$),  the larger $w_i$, the larger the amplitude of $\delta_{\rm EDE}$ and therefore the larger the contribution to the Weyl potential around $a_c$. For the mode entering the horizon around $a_c$ ($k=0.06$Mpc$^{-1}$) one can note again the non-trivial behavior, where the model with smaller $w_i$ shows a larger first oscillation. This is due to the fact that $w$ has evolved in time, affecting the damping term, as discussed previously.  For the modes that enter the horizon very late ($k=0.001$Mpc$^{-1}$), the EDE perturbations are essentially zero, and the weyl potential is suppressed around $a_c$ due to the presence of a non-clustering EDE, with minute differences due to slightly-different time evolution in $w_{\rm EDE}$. 
\end{itemize}

\subsection{Impact of $w_i$ on the CMB and matter power spectra}

In figure \ref{ede_cl_pk} left panel, we plot residuals of the CMB temperature anisotropy power spectra  (TT) when varying $w_i$ and keeping other parameters fixed, taking $\Lambda$CDM as reference. The main effect of EDE is described in detail in Ref.~\cite{Poulin:2023lkg}. Here we focus on describing the impact of varying $w_i$. The main visible impact of $w_i$ on the CMB TT power spectra comes through the following contributions
\begin{itemize}
    \item Diffusion damping: When we increase $w_i$, the effect of EDE on the background expansion is increased. Because we keep $\theta_s$ fixed, and the impacts of EDE on the sound horizon and on the damping scales are different, the adjustment of the angular diameter distance $D_A$ by the increase in $H_0$ cannot simultaneously keep the angular diffusion damping scale $\theta_{d}=\frac{r_{d}}{D_A}$ unaffected. As a result, $\theta_{d}$ increases, leading to a suppression of CMB TT power spectra at high $\ell$ that is larger for larger $w_i$. 
    
     \item Sachs-wolfs contribution: As a result of the changes to the Weyl potential due to increasing $w_i$ (and the longer lasting EDE phase), the Sachs wolf contribution is significantly affected around $\ell \sim 500$. The larger $w_i$, the larger the contribution in the Sachs wolf effect at intermediate $\ell$'s, boosting the first acoustic peak amplitudes.

\end{itemize}

 Next, in figure \ref{ede_cl_pk} right panel, we show the effect of increasing $w_i$ on the total matter power spectra. When we increase, $w_i$ the matter power at smaller scales is suppressed compared to $w_i=-1$ case, due to the longer period of EDE (that suppresses the Weyl potential). As a result, we get a lower $\sigma_{8}$ for larger $w_i$, showing that letting $w_i$ free to vary may help in alleviating the tension with weak lensing surveys. The increase of power at larger scales is due to the fact that  $\omega_\Lambda = 1-\Omega_{\rm m}$ is smaller in models with larger $w_i$, as a by product of the larger $h$ value that has increased $\Omega_{\rm m}  = \omega_{\rm m}h^2$ at fix $\omega_{\rm m}$.

\section{Details of the analysis}
\subsection{Data sets}\label{sec:datasets}
We use a combination of CMB and BAO data sets along with S$H_0$ES and KIDS priors. The details of our data sets are as follows:
\begin{itemize}
    \item Planck 2018 measurements of the low-$\ell$ CMB TT, EE, and  high-$\ell$ TT, TE, EE power spectra, together with the gravitational lensing potential reconstruction \cite{Aghanim:2018eyx}. 
     \item The BAO measurements from 6dFGS at $z=0.106$~\cite{Beutler:2011hx}, SDSS DR7 at $z=0.15$~\cite{Ross:2014qpa}, BOSS DR12 at $z=0.38, 0.51$ and $0.61$~\cite{Alam:2016hwk}, and the joint constraints from eBOSS DR14 Ly-$\alpha$ auto-correlation at $z=2.34$~\cite{Agathe:2019vsu} and cross-correlation at $z=2.35$~\cite{Blomqvist:2019rah}.
    \item The measurements of the growth function $f\sigma_8(z)$ (FS) from the CMASS and LOWZ galaxy samples of BOSS DR12 at $z = 0.38$, $0.51$, and $0.61$~\cite{Alam:2016hwk}.
    \item The Pantheon SNIa catalogue, spanning redshifts $0.01 < z < 2.3$~\cite{Scolnic:2017caz}. We anticipate that the new Pantheon+ \cite{Brout:2022vxf} would not significantly affect our conclusions.
     \item  The S$H_0$ES result, modeled with a Gaussian likelihood centered on $H_0 = 73.2 \pm 1.3$ km/s/Mpc \cite{Riess:2020fzl}; however, choosing a different value that combines various direct measurements, or the updated value from \cite{Riess:2021jrx} would not significantly affect our conclusions. 
     \item The KIDS1000+BOSS+2dfLenS weak lensing data, compressed as a split-normal likelihood on the parameter $S_8=0.766^{+0.02}_{-0.014}$~\cite{Heymans:2020gsg}. We note that there are additional measurements from DES \cite{DES:2021wwk} and the combination o KiDS and DES \cite{Kilo-DegreeSurvey:2023gfr} that show lower tension with {\it Planck} under $\Lambda$CDM, but we choose this value as a representative example. Choosing a different prior would not drastically change our conclusions.
\end{itemize}

\subsection{Methodology}
Our baseline cosmology consists of the following combination of the six $\Lambda$CDM parameters $\{\omega_b,\omega_{\rm cdm},100\times\theta_s,n_s,{\rm ln}(10^{10}A_s),\tau_{\rm reio}\}$, plus 
four EDE parameters as discussed in Sec~\ref{sec:background}, namely {$ w_i$, $w_f$, $z_{\rm c}$, $f_{\rm EDE}$}. 
We run MCMC analyses of the EDE model against various combinations of the CMB, BAO, and supernovae data sets (details of which are given in Sec~\ref{sec:datasets})
with the Metropolis-Hasting algorithm as implemented in the MontePython-v3 \cite{Brinckmann:2018cvx} code interfaced with our modified version of CLASS. All reported $\chi^2_{\rm min}$ are obtained with the python package {\sc iMinuit \footnote{\url{https://iminuit.readthedocs.io/}}} \cite{James:1975dr}. We make use of a Choleski decomposition to better handle a large number of nuisance parameters \cite{Lewis:1999bs} and consider chains to be converged with the Gelman-Rubin convergence criterion $R-1<0.05$ \cite{Gelman:1992zz}.

We perform analyses of three different variations of the EDE model: i) a two-parameter fluid model of EDE  varying only ($f_{\rm EDE}$,$\rm z_c$) while fixing the equation of state parameters as ($w_i=-1$,$w_{f}=1$), that we dub 2pEDE; ii) a  three-parameter ($w_f$,$f_{\rm EDE}$,$z_c$) model dubbed 3pEDE; iii) a four-parameter ($ w_i$,$ w_f$,$f_{\rm  EDE}$,$z_c$) dubbed 4pEDE. For each model, we perform three sets of runs, starting from the baseline Planck+BAO+Pantheon, then adding the S$H_0$ES prior, and finally the $S_8$ prior. We also perform the same sets of runs with $\Lambda$CDM for comparison purposes.  We set large flat prior for all $\Lambda$CDM parameters. Prior ranges for Early dark energy parameters are imposed as follows:\\

\begin{table}[h!]
\centering
\begin{tabular}{|c|c|}
\hline
   Parameter name  &  prior range\\
   \hline
    $w_i$ &[-1,0]\\
    $w_f$&[0,1]\\
    $f_{\rm EDE}$& [0,0.3]\\
    $\log_{10}(z_c)$&[2,5]\\
    \hline
\end{tabular}
\end{table}
Note that we let $w_i$ be greater than $-1/3$, such that EDE does not formally refer to a DE like component in some part of the parameter space. Nevertheless, there is nothing that becomes mathematically ill defined in the equations. We explore the $w_i > 0 $ part of the parameter space in App.~\ref{app:prior}.
\section{Results}
\label{sec:results}
\subsection{Results including the S$H_0$ES prior}

\begin{table*}
    \centering
    \resizebox{\textwidth}{!}{
    \begin{tabular}{|l|c|c|c|c|}
    \hline
  Parameters $\downarrow$ &$\Lambda CDM$ &2 param EDE &3 param EDE &  4 param EDE \\
     \hline 
     
     $100~\theta_s$&$1.042066(1.04198)_{-0.00028}^{+0.00029}$&$1.04158(1.04159)_{-0.00035}^{+0.00034}$&$1.04160(1.04156)_{-0.00038}^{+0.00033}$ &$1.04124(1.04094)_{-0.00053}^{+0.0006}$\\
     $100~\omega_{b}$&$2.253(2.249)_{-0.014}^{+0.013}$&$2.284(2.283)_{-0.023}^{+0.022}$&$2.283(2.278)_{-0.025}^{+0.022}$&$2.285(2.271)_{-0.024}^{+0.023}$\\
     $\omega_{\rm cdm}$&$0.1184(0.1183)_{-0.00088}^{+0.00089}$&$0.1258(0.1247)_{-0.0031}^{+0.0035}$&$0.1274(0.1267)_{-0.0032}^{+0.0034}$ &$0.128(0.1326)_{-0.0041}^{+0.0044}$\\
     $\log{10^{10} A_{s}}$&$3.054(3.056)_{-0.016}^{+0.015}$ &$3.063(3.067)_{-0.015}^{+0.015}$ &$3.065(3.072)_{-0.017}^{+0.015}$ &$3.063(3.066)_{-0.016}^{+0.015}$\\
     $n_s$& $0.9697(0.9701)_{-0.0037}^{+0.0037}$&$0.9803(0.9819)_{-0.0062}^{+0.0062}$ &$0.9841(0.9849)_{-0.0068}^{+0.0066}$ &$0.983(0.9922)_{-0.007}^{+0.0078}$\\
     $\tau_{\rm reio}$&$0.0602(0.0617)_{-0.0082}^{+0.0073}$&$0.0569(0.0600)_{-0.0076}^{+0.007}$ &$0.0573(0.0616)_{-0.0078}^{+0.0073}$ &$0.0578(0.0554)_{-0.0081}^{+0.0071}$ \\
  
     \hline
     $f_{\rm EDE}$ &$-$&$0.112(0.105)_{-0.036}^{+0.047}$ &$0.118(0.121)_{-0.034}^{+0.044}$ & $0.112(0.160)_{-0.04}^{+0.054}$ \\
    ${\log_{10}(z_c)}$&$-$&$3.53(3.51)_{-0.12}^{+0.09}$ &$3.67(3.61)_{-0.15}^{+0.091}$ &$3.82(3.81)_{-0.23}^{+0.15}$ \\
    $w_i$ &$-$ & $-$ &$-$ & $-0.651(-0.783)_{-0.35}^{+0.086}$\\
    $w_f$& $-$&$-$ &$0.74(0.79)_{-0.13}^{+0.12}$ &$0.61(0.60)_{-0.13}^{+0.1}$\\
  
    \hline

    $\sigma_{8}$ & $0.8097(0.8108)_{-0.0064}^{+0.006}$&$0.831(0.8305)_{-0.011}^{+0.011}$ & $0.834(0.836)_{-0.011}^{+0.011}$&$0.830(0.841)_{-0.011}^{+0.012}$  \\
    $H_0$ [km/s/Mpc] &$68.18(68.12)_{-0.41}^{+0.39}$ &$70.03(70.09)_{-0.85}^{+0.91}$ &$70.46(70.52)_{-0.91}^{+0.9}$ &$70.68(71.59)_{-1.1}^{+1.2}$\\
    \hline 
    $\chi^2_{\rm min}$ &3826.58 &3816.46&3814.53 &3812.16\\
    \hline 
    $\Delta \chi^2_{\rm min}$ & 0&-10.12& -12.05 &-14.5 \\
    \hline 
    \end{tabular}}
    \caption{The mean (best-fit) $\pm1\sigma$ error of the cosmological parameters reconstructed from the lensing-marginalized Planck+BAO+SN1a data with H0 prior }
    \label{tab:MCMC_H0}
\end{table*}

\begin{figure*}[t!]
    \centering
\includegraphics[scale=0.4]{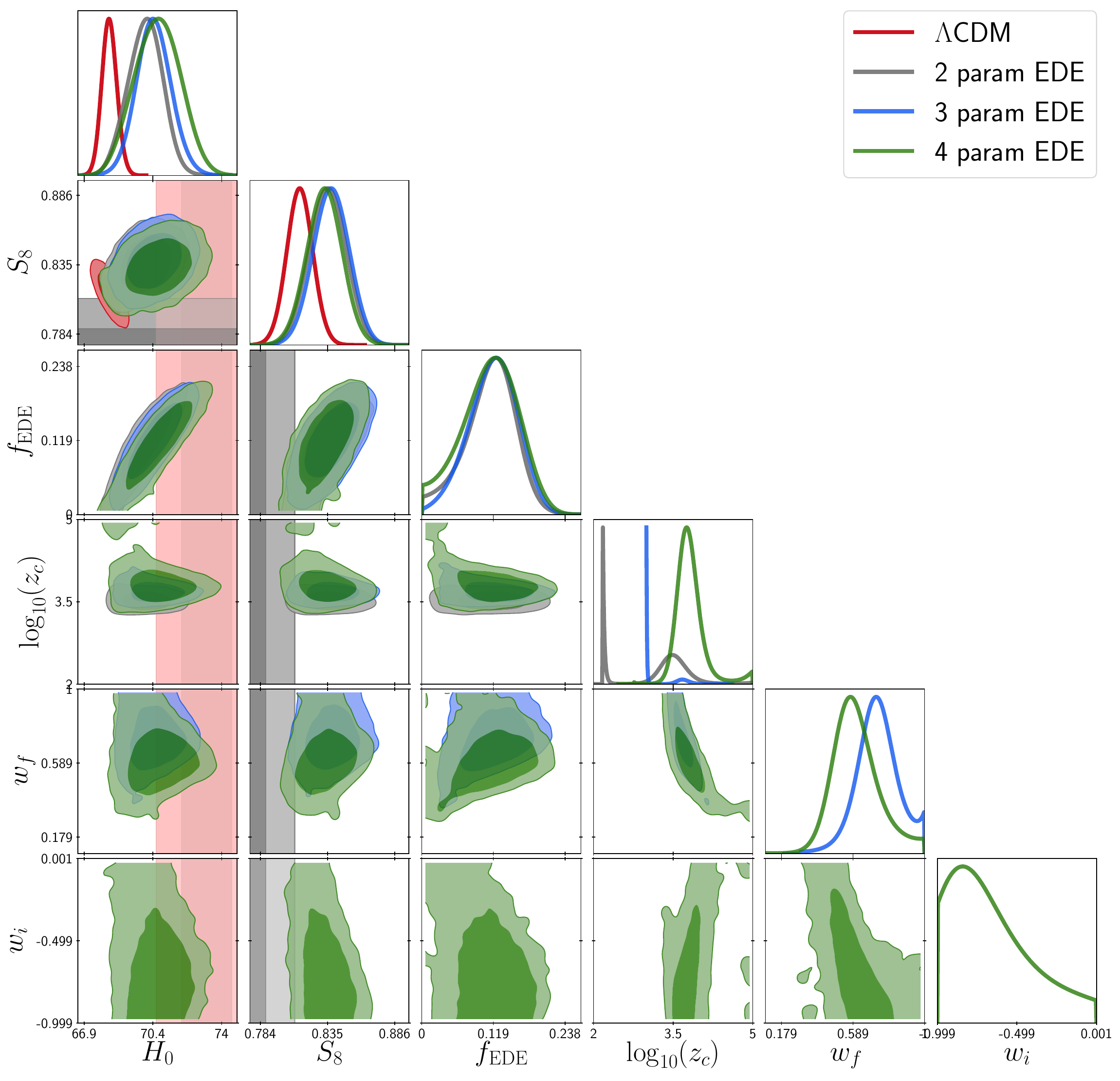}
    \caption{Posterior distributions in the $\Lambda$CDM and EDE models reconstructed from Planck+BAO+Pantheon+$H_0$.}
    \label{fig:H0}
\end{figure*}

We start by comparing the ability of the different models to resolve the Hubble tension and therefore focus on the Planck+BAO+Pantheon+S$H_0$ES analyses.
The reconstructed mean and best-fit values of parameters are given in table \ref{tab:MCMC_H0}. We provide $\chi^2_{\rm min}$ for $\Lambda$CDM vs 2pEDE vs  3pEDE vs 4pEDE model  in App.~\ref{app:chi2} in table \ref{tab:chi2_ede3} (without S$H_0$ES) and table \ref{tab:chi2_ede1} (with S$H_0$ES). We plot the 1D and 2D posterior distribution of the EDE parameters as well as $H_0$ and $S_8$ in figure \ref{fig:H0} for $\Lambda$CDM compared to the three different EDE models.

From the results, it is evident that the value of the Hubble parameter is higher in the case of the 4pEDE model ($H_0=70.68_{-1.1}^{+1.2}$) compared to the 3pEDE model ($H_0=70.46_{-0.91}^{+0.9}$) and 2pEDE model ($H_0=70.03_{-0.85}^{+0.91}$). 
In fact, the overall $\chi^2_{\rm min}$ is also improved by 2.4 in the 4pEDE model compared to the 3pEDE model, which is a slightly larger improvement in $\chi^2$ than when going from 2pEDE to 3pEDE. However, the value of $w_i$ is only weakly constrained, with only an upper limit at $1\sigma$ of $w_i < -0.565$, but compatible with $0$ at $2\sigma$. In App.~\ref{app:prior}, we find that the only hard limit on $w_i$ is $w_i < 1/3$, from the requirement that EDE does not dominate the energy density at early times. 

Most importantly, while the value of $H_0$ is slightly larger,  the value of  $\sigma_8$ has slightly decreased, by $\sim 0.4\sigma$. This indicates that $w_i$ may indeed play a role in the $S_8$ tension, and we now turn to include $S_8$ measurements in the analysis.

\begin{table*}
    \centering
    \begin{tabular}{|l|c|c|c|c|}
    \hline
  Parameters $\downarrow$ &$\Lambda CDM$ & 3 param EDE &  4 param EDE \\
     \hline 
     
     $100~\theta_s$&$1.04210(1.04214)_{-0.00028}^{+0.00028}$ & $1.04185(1.04178)_{-0.00042}^{+0.00034}$ &$1.04143(1.04163)_{-0.00051}^{+0.00072}$\\
     $100~\omega_{b}$&$2.258(2.268)_{-0.013}^{+0.013}$&$2.277(2.271)_{-0.023}^{+0.02}$&$2.281(2.271)_{-0.024}^{+0.021}$\\
     $\omega_{\rm cdm}$&$0.1177(0.1179)_{-0.00081}^{+0.00085}$&$0.1227(0.1225)_{-0.0035}^{+0.0029}$ &$0.1238(0.1228)_{-0.0044}^{+0.0034}$\\
     $\log{10^{10} A_{s}}$& $3.048(3.037)_{-0.015}^{+0.014}$&$3.053(3.050)_{-0.015}^{+0.015}$ & $3.053(3.044)_{-0.015}^{+0.015}$\\
     $n_s$&$0.9708(0.9712)_{-0.0037}^{+0.0036}$  &$0.9797(0.9814)_{-0.0074}^{+0.0072}$ &$0.9795(0.9802)_{-0.0074}^{+0.0071}$\\
     $\tau_{\rm reio}$&$0.0580(0.0520)_{-0.0078}^{+0.007}$&$0.0562(0.0544)_{-0.0074}^{+0.0072}$ & $0.0569(0.0526)_{-0.0077}^{+0.0071}$\\
  
     \hline
     $f_{\rm EDE}$ &$-$& $0.069(0.068)_{-0.048}^{+0.036}$ &$0.075(0.054)_{-0.048}^{+0.041}$\\
    ${\log_{10}(z_c)}$&$-$& $3.79(3.72)_{-0.32}^{+0.14}$ &$3.89(4.02)_{-0.33}^{+0.22}$\\
    $w_i$ &$-$ & $-$ & unconstrained (-0.34)\\
    $w_f$& $-$ & unconstrained$(0.65)$ &$0.58(0.45)_{-0.16}^{+0.1}$\\
  
    \hline

    $\sigma_{8}$ &$0.8051(0.8015)_{-0.006}^{+0.0057}$ & $0.817(0.817)_{-0.011}^{+0.01}$ & $0.8161(0.8109)_{-0.0096}^{+0.0095}$  \\
    $H_0$ [km/s/Mpc] & $68.48(68.50)_{-0.38}^{+0.38}$&$69.94(69.92)_{-1}^{+0.92}$ &$70.25(69.93)_{-1.2}^{+1.1}$\\
    \hline 
    $\chi^2_{\rm min}$ &3832.41 &3823.99  &3823.79\\
    \hline 
    $\Delta \chi^2_{\rm min}$ & 0&-8.42  &-8.62 \\
    \hline 
    \end{tabular}
    \caption{The mean (best-fit) $\pm1\sigma$ error of the cosmological parameters reconstructed from Planck+BAO+SN1a data with $H_0+S_8$ priors. }
    \label{tab:MCMC_S8}
\end{table*}

\subsection{Results including the $S_8$ prior}

\begin{figure*}[t!]
    \centering
   
    \includegraphics[scale=0.4]{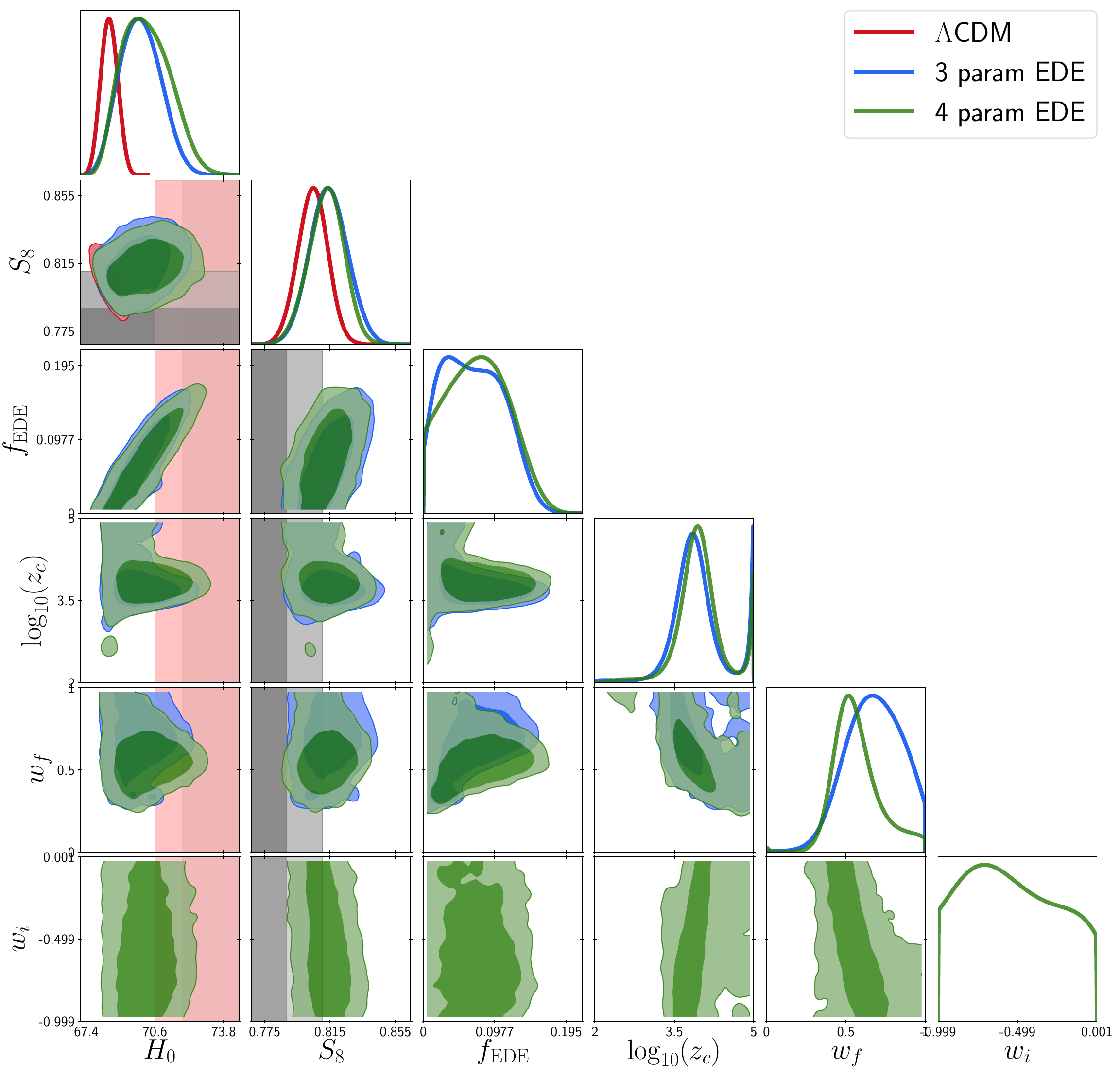}
    \caption{Same as Fig.~\ref{fig:H0}, now also include the $S_8$ prior.}
    \label{fig:S8}
\end{figure*}

The mean and best-fit values of various parameters are reported in table \ref{tab:MCMC_S8} and we plot the same parameters as before in Figure \ref{fig:S8}. All  $\chi^2_{\rm min}$ numbers are provided in table \ref{tab:chi2_ede2}.

The main impact of adding the $S_8$ prior is to reduce the preference for non-zero EDE and decrease the value of $H_0$ while pulling $\sigma_8$ down. The $\Delta\chi^2$ with respect to $\Lambda$CDM is also significantly decreased compared to the case without $S_8$ prior.  In addition, the 4pEDE model fit is only marginally better than the 3p EDE model, and $w_i$ is now unconstrained. We conclude that, even though the addition of $w_i$ does slightly decrease the $S_8$ parameter, it cannot help in resolving both $H_0$ and $S_8$ tension simultaneously  in the EDE model.

\section{Conclusions}
\label{sec:concl}
We have explored the cosmological implications of early dark energy beyond slow roll (e.g. non-cc equation of state), by changing the initial equation of state $w_i$, usually set to $-1$. Our main findings are as follows:

\begin{itemize}
    \item  When one increases $w_i$ at fix $f_{\rm EDE}$ and $z_c$, EDE contributes for a longer time to the expansion rate prior to recombination, resulting in a smaller sound horizon of the CMB and hence a larger $H_0$.

    \item The background effect leads to a suppression of the Weyl potential, due to the larger contribution of non-clustering EDE to the total energy density.  This results in a comparatively lower power at small scales and hence a smaller $\sigma_8$. We also find that there are additional perturbative effects due to $w_i$ for modes that enter the horizon before or around $z_c$, although the effects on observable are small compared to the main background effect. 

    \item We have confronted three variants of the EDE model to the combination of Planck18+BAO+Pantheon+S$H_0$ES: a two-parameter EDE model with $w_i$ and $w_f$ fixed, a three-parameter EDE model with $w_f$ freed, and the four-parameter EDE model with both $w_i$ and $w_f$ free. We have found that the overall $\chi^2_{\rm min}$ is improved by -2.4 at the expense of one more parameter $w_i$ compared to the 3-parameters model,  and by -4.4 compared to the 2-parameters  one. However, $w_i$ is not detected in this analysis, and we only derive a weak upper limit at $1\sigma$, $w_i < -0.565$.
    
   \item Interestingly, the model with $w_i$ free has a slightly larger $H_0$, and a value of $S_8$ decreased by $\sim 0.4 \sigma$. However, the inclusion of $S_8$ data reduces the preference for non-zero EDE, with a degradation in the $\chi^2_{\rm min}$.
\end{itemize}

Although the results derived in this work indicate that a non-cc EDE cannot resolve both $H_0$ and $S_8$ tensions simultaneously, we hope that it will trigger further work towards mitigating the increase of small-scale power that is induced in the EDE cosmology. In fact, this is a requirement not only of EDE, but generally of the fact that a larger $H_0$ (as measured by S$H_0$ES) and a well-constrained $\Omega_m$ (as measured by Pantheon and BAO data) must imply a larger $\omega_m\equiv \Omega_m h^2$, and therefore earlier matter domination and larger $\sigma_8$. Models of EDE in that sense already manage to compensate for the effect of the larger $H_0$ and the larger $\omega_m$ to adjust CMB data but do not sufficiently reduce the growth of structures. It therefore remains to explore how one may further build upon these results, within an EDE cosmology \cite{McDonough:2021pdg} or others \cite{Joseph:2022jsf,Schoneberg:2023rnx}, to finally resolve cosmic tensions simultaneously.
\section*{Acknowledgements}
We acknowledge HPC NOVA, IIA Bangalore where numerical
simulations were performed. 
VP is supported by funding from the European Research Council (ERC) under the
European Union’s HORIZON-ERC-2022 (Grant agreement No.~101076865).

\pagebreak

\bibliography{output}{}
\bibliographystyle{aasjournal}


\appendix

\section{Effect of changing prior on $w_i$}
\label{app:prior}
\begin{figure*}
    \centering
    \includegraphics[scale=0.4]{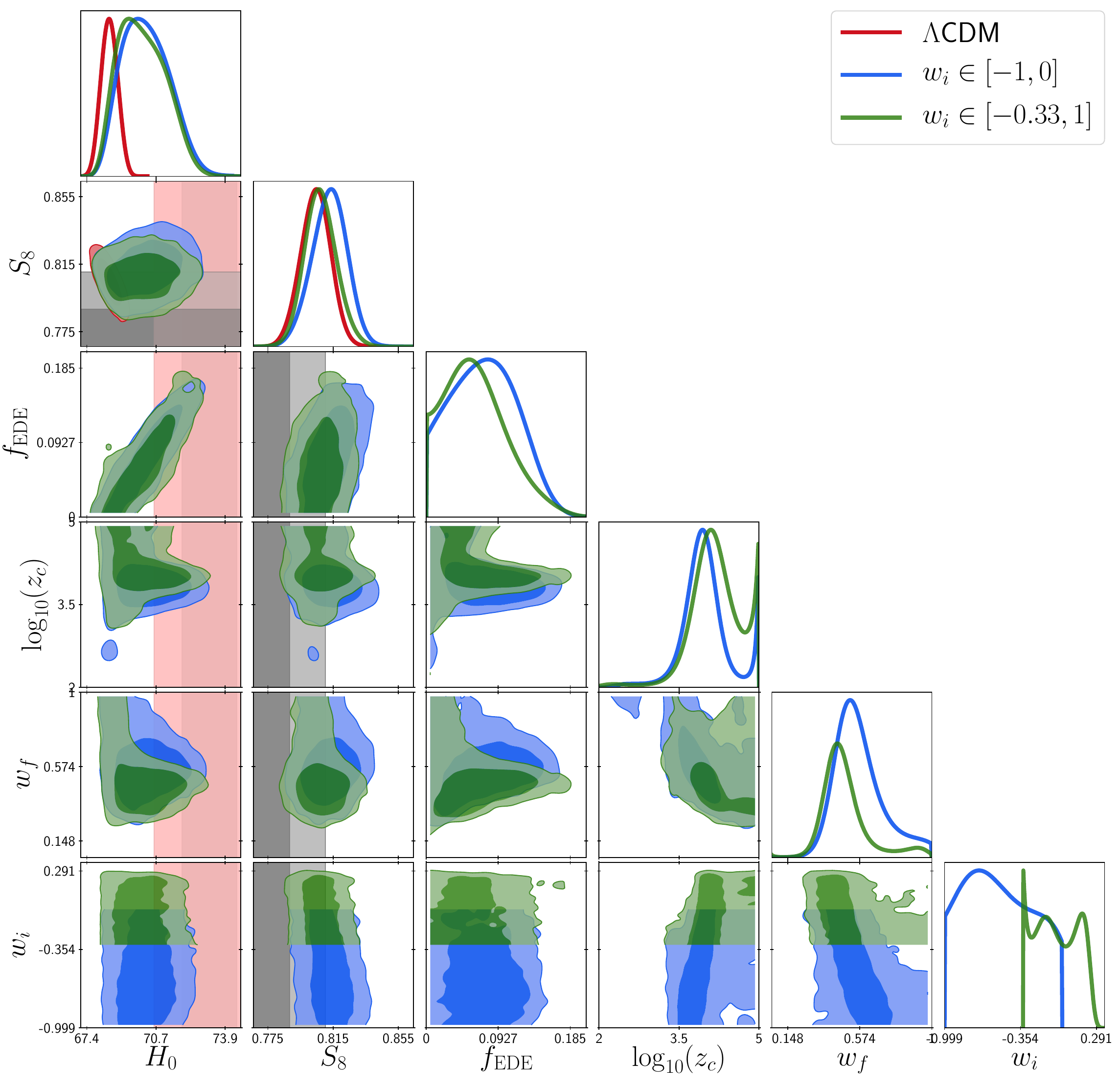}
    \caption{Effect of changing prior on $w_i$}
    \label{fig:prior_wi}
\end{figure*}

In the main text, we have examined the possibility of changing the initial equation of state parameter $w_i \in  [-1,0]$. Now let us examine the case beyond Dark Energy, with $w_i \in [-0.33,1]$. The rest of the parameters are varied as in the main body of the paper. Our main results are shown in figure \ref{fig:prior_wi}, which compare the results of $w_i \in [-0.33,1]$ with the regular prior given in the main text, when Planck+BAO+Pantheon+$H_0$+$S_8$ data  are used. We also provide in table \ref{tab:MCMC_prior} the parameters reconstructed either when including only the $H_0$ prior or both the $H_0$ and $S_8$ priors. 

First, the posterior of $w_i$ has a strict upper bound around 0.3, which simply comes from the fact that a fluid with larger $w_i$ would come to dominate the expansion rate at early times and spoil the fit to data. Second, one can see that the posterior of $S_8$ exactly matches with $\Lambda$CDM, while $H_0$ is significantly larger. In other words, an ``EDE''  model with $w_i > -0.33$ can perform as good as regular EDE in resolving the tension without worsening the $S_8$ tension. 
In terms of $\chi^2$ number, the model of EDE with $w_i \in [-1,0]$ performs slightly better (by $-1.2$) when the $H_0$ prior is left out of the analysis. However, the model with $w_i \in [-1/3, 0]$ performs  better when $S_8$ is included, as the EDE contribution lasts even longer, further reducing the growth of the structure.

\begin{table*}
    \centering
    \begin{tabular}{|l|c|c|c|}
    \hline
  Parameters $\downarrow$ &  $H_0$ prior & $H_0$+$S_8$ prior  \\
     \hline 
     
     $100~\theta_s$&$1.04143(1.04036)_{-0.00095}^{+0.0011}$&$1.04108(1.04026)_{-0.00065}^{+0.0013}$\\
     $100~\omega_{b}$& $2.278(2.285)_{-0.024}^{+0.02}$&$2.279(2.272)_{-0.023}^{+0.019}$\\
     $\omega_{\rm cdm}$& $0.126(0.1298)_{-0.0042}^{+0.0042}$&$0.1233(1.268)_{-0.0051}^{+0.0027}$\\
     $\log{10^{10} A_{s}}$& $3.059(3.060)_{-0.015}^{+0.015}$&$3.051(3.062)_{-0.016}^{+0.014}$\\
     $n_s$& $0.979(0.9826)_{-0.0067}^{+0.007}$&$0.977(0.9810)_{-0.0075}^{+0.0061}$\\
     $\tau_{\rm reio}$& $0.0590(0.0576)_{-0.0078}^{+0.0073}$&$0.0577(0.0605)_{-0.0079}^{+0.0069}$\\
  
     \hline
     $f_{\rm EDE}$ & $0.081(0.120)_{-0.045}^{+0.041}$&$0.065(0.094)_{-0.055}^{+0.028}$\\
    ${\log_{10}(z_c)}$& $4.11(3.96)_{-0.28}^{+0.12}$&$4.134(4.038)_{-0.5}^{+0.31}$\\
    $w_i$ & $<0.33 (-0.20)$ &$<0.33 (0.0978)$\\
    $w_f$& $0.461(0.483)_{-0.089}^{+0.082}$&$0.486(0.409)_{-0.15}^{+0.057}$\\
  
    \hline 
      $\sigma_{8}$ & $0.8195(0.8252)_{-0.0087}^{+0.0085}$   &$0.8109(0.8172)_{-0.0088}^{+0.0074}$\\
    $H_0$ [km/s/Mpc] & $70.23(70.97)_{-1.1}^{+1.1}$&$70.08(70.78)_{-1.3}^{+0.99}$\\
    \hline 
    $\chi^2_{\rm min}$&3813.32 &3822.69\\
    \hline 
    $\Delta \chi^2_{\rm min}$&-13.26 &-9.72 \\
    \hline 
    \end{tabular}
    \caption{The mean (best-fit) $\pm1\sigma$ error of the cosmological parameters reconstructed from Planck+BAO+SN1a data with H0 and $S_8$ when $w_i \in [-0.333,1]$ }
    \label{tab:MCMC_prior}
\end{table*}

\section{Tables of $\chi^2$}\label{app:chi2}

\begin{table*}
\scalebox{0.9}{
  \begin{tabular}{|l|c|c|c|c|c|c|}
  \hline
   \multicolumn{1}{|c|}{Experiments/Data}&\multicolumn{1}{|c|}{$\Lambda$CDM}&\multicolumn{1}{|c|}{2-param EDE}&\multicolumn{1}{|c|}{3-param EDE}&\multicolumn{1}{|c|}{4-param EDE } \\
  \hline
  Planck~high$-\ell$ TT,TE,EE &2347.90&2348.99&2346.49& 2346.95 \\
Planck~ low$-\ell$ EE &396.60&396.03 &396.65&396.63  \\
Planck~ low$-\ell$ TT &22.97&22.22&21.89& 22.19  \\
Planck~lensing &8.76&9.016&9.04&9.16 \\
Pantheon &1025.92 &1025.99&1025.81&1026.02\\
BAO~FS~BOSS DR12 & 6.61 &7.23&6.68&7.38\\
BAO~BOSS low$-z$ &1.19&1.10&1.30&1.06\\

\hline
total &3809.99 &3810.59&3807.90&3809.44 \\
\hline
  \end{tabular}  
  }
  \caption{Best-fit $\chi^2$ per experiment (and total) when no prior was included.}
  \label{tab:chi2_ede3}
\end{table*}

\begin{table*}
\scalebox{0.9}{
  \begin{tabular}{|l|c|c|c|c|c|c|c|c|}
  \hline
   \multicolumn{1}{|c|}{Experiments/Data}&\multicolumn{1}{|c|}{$\Lambda$CDM}&\multicolumn{1}{|c|}{2-param EDE }&\multicolumn{1}{|c|}{3-param EDE}&\multicolumn{1}{|c|}{4-param EDE}&\multicolumn{1}{|c|}{4-param $w_i \in [-0.333,1]$} \\
  \hline
  Planck~high$-\ell$ TT,TE,EE &2348.49&2349.17&2347.96&2349.86&2349.51   \\
Planck~ low$-\ell$ EE &397.97&397.16& 397.77&396.09&396.63  \\
Planck~ low$-\ell$ TT &22.67& 21.57&21.33&20.69&21.37   \\
Planck~lensing &8.95&8.96&9.16&9.81&9.24 \\
Pantheon & 1025.69&1025.64&1025.64&1025.70&1025.68 \\
BAO~FS~BOSS DR12 &5.93 &6.42&6.61&6.99&6.42\\
BAO~BOSS low$-z$ &1.61&1.82&1.79&1.47&1.51\\
S$H_0$ES &15.23&5.68&4.23&1.52&2.92   \\ 

\hline
total & 3826.58& 3816.46&3814.53&3812.16&3813.32 \\
\hline
  \end{tabular}  
  }
  \caption{Best-fit $\chi^2$ per experiment (and total) when the S$H_0$ES prior is included.}
  \label{tab:chi2_ede1}
\end{table*}

\begin{table*}
\scalebox{0.9}{
  \begin{tabular}{|l|c|c|c|c|c|c|}
  \hline
   \multicolumn{1}{|c|}{Experiments/Data}&\multicolumn{1}{|c|}{$\Lambda$CDM}&\multicolumn{1}{|c|}{3-param EDE}&\multicolumn{1}{|c|}{4-param EDE}&\multicolumn{1}{|c|}{4-param $w_i \in [-0.333,1]$} \\
  \hline
  Planck~high$-\ell$ TT,TE,EE &2353.87&2350.49&2351.38 &2350.96  \\
Planck~ low$-\ell$ EE &395.68& 395.89&395.75&397.42  \\
Planck~ low$-\ell$ TT &22.30&21.29&21.32&21.56   \\
Planck~lensing &10.63&10.02&10.46&9.40 \\
Pantheon & 1025.63&1025.62&1025.63&1025.62 \\
BAO~FS~BOSS DR12 & 5.87 &6.29&6.10&6.14\\
BAO~BOSS low$-z$ &1.92&2.154&2.069&1.90\\
S$H_0$ES &13.06&6.36&6.29&3.40   \\ 
$\rm S_8(KIDS1000)$&3.41&5.84&4.74&6.25\\
\hline
total & 3832.41&3823.99&3823.79&3822.69 \\
\hline
  \end{tabular}  
  }
  \caption{Best-fit $\chi^2$ per experiment (and total) when S$H_0$ES$+S_8$ prior is included.}
  \label{tab:chi2_ede2}
\end{table*}

\end{document}